\documentstyle[aps,prb,epsfig]{revtex}
\begin{document}
\draft

\title{Improved stability regions for ground states of the extended Hubbard
  model}
\author{Zsolt~Szab\'o \cite{cim}}
\address{Institut f\"ur Theoretische Physik, Universit\"at zu K\"oln,
Z\"ulpicher Str. 77, D-50937 K\"oln, Germany}
\date{31 December 1998}

\maketitle
\begin{abstract}
The ground state phase diagram of the extended Hubbard model containing 
nearest and next-to-nearest neighbor interactions is investigated in the 
thermodynamic limit using an exact method. It is found that taking into 
account local correlations and adding next-to-nearest neighbor
interactions both have significant effects on
the position of the phase boundaries. Improved stability domains for the
$\eta$-pairing state and for the fully saturated ferromagnetic state at half
filling have been constructed. The results show that these states are the 
ground states for model Hamiltonians with realistic values of the interaction
parameters.\\

\end{abstract}

\section{Introduction}
\label{sec:intro}

Exact solutions in physics are of great importance, since in some cases the 
errors introduced by the approximations may dominate the results to 
such an extent that one might end up with an incorrect description 
of the studied phenomenon. When applying an analytic, but non-exact, approach 
one has to know to what extent the 
approximation is valid. In case of perturbative methods a well-defined 
{\em small} parameter can ensure that the higher order terms are indeed 
negligible. Often, it is hard to find such a small parameter due to the fact 
that the given phenomenon itself has strong-coupling characteristics,
i.e. the associated correlation effects are not small at all. This is
especially true by the investigation of strongly correlated electron systems. 
Ferromagnetism is an example for {\em intermediate-to-strong} coupling 
phenomenon, where one has to be very cautious to apply perturbative
approaches. 

Considering the exact results with respect to the dimensionality {\em D}, 
we can see that most of them have been derived in two limiting cases: either 
 $D\!=\!1$ or {\em $D\!=\!\infty$}. For example, the exact solution of
the Hubbard 
model was given in $D\!=\!1$ dimension by means of the Bethe-ansatz by Lieb 
and Wu \cite{lieb}. The other class of exact solutions belongs to the other 
limiting case, i.e. {\em $D\!=\!\infty$}, when the dynamical mean-field
approximation becomes exact \cite{metzner,muller-hartmann}. 
The situation, however, gets more complicated as physically 
interesting, lower dimensional cases (e.g. systems in $D\!=\!2$ or $D\!=\!3$ 
dimensions) are considered. $D \!>\! 1$ rules out the applicability 
of the well-established Bethe-ansatz approach, while mean-field like
descriptions lead to qualitatively or quantitatively incorrect conclusions,
because the effects of spatial fluctuations are not taken properly into 
account \cite{amadon,ulmke}.   

In recent years, some exact, {\em non-perturbative} methods have 
been developed to investigate the ground state of Hubbard and 
Hubbard-like models in large parameter regimes
\cite{essler,ovhinnikov2,brandt,mielke,strack1,strack2,boer1,hanisch}. 
In the present work we focus on the so-called optimum ground state method 
(OGS) established by de Boer and Schadschneider \cite{boer1}. Using this 
method one can obtain rigorous constraints on the model parameters which
define regions where e.g. the CDW, the N\'eel, the fully saturated 
ferromagnetic, or the $\eta$-pairing state of momentum {\em P} becomes 
the exact ground state of the Hamiltonian. 

The basic idea of the OGS method is to diagonalize a specially chosen 
local Hamiltonian and to tune all the model-parameters such a way that all 
local eigenstates which are needed to the construction of a given global 
ground state are also local ground states. This means that, on one hand, 
the corresponding eigenvalues of the local Hamiltonians should be all equal 
in magnitude and, on the other hand, this common value should be the lowest 
eigenvalue of the local problem. Following this method, one can obtain 
different regions in the parameter space of the model defined by inequalities. 
These inequalities mean {\em sufficient} conditions for a state
to be  the ground state inside a special region. Outside the derived
region the state under study may or may not be the ground state of the model. 

There are basically two different ways to enlarge the region of guaranteed
stability: taking a larger local Hamiltonian or incorporating 
next-to-nearest neighbor interactions. As far as the first approach is
concerned, it is obvious that the extent of local correlations which
are taken into account is controlled by the size of the local
Hamiltonian to be diagonalized exactly. Therefore, using local
Hamiltonians defined on larger clusters of the lattice should
tipically lead to better constraints (i.e. extended stability
regions), even if purely nearest neighbor interactions are present.  

It is also well-known that next-to-nearest neighbor hopping has important 
effects. For instance, it was shown rigorously by Tasaki \cite{tasaki} 
that the pure Hubbard model characterized by hopping of electrons between 
nearest and next-to-nearest neighboring sites with dispersive bands exhibits
ferromagnetism for finite Coulomb interaction at zero temperature.
Recent Projection Quantum Monte Carlo studies by Hlubina {\em et al.} 
\cite{hlubina} confirmed this fact for finite temperatures, too. Beside 
the consequences of longer range hopping, the importance of nearest and 
next-to-nearest neighbor off-site interactions (both diagonal and
off-diagonal) has also been emphasized both from experimental \cite{verdozzi} 
and from theoretical 
\cite{amadon,strack1,strack2,brink,szabo} sides. These extra terms 
(density-density type interaction, correlated hopping of electrons, hopping of
electron pairs and the exchange coupling) all originate from the spin
independent Coulomb interaction of electrons. Nevertheless, the
values of these longer range interactions (e.g. between next-to-nearest 
neighboring sites) are known neither theoretically nor experimentally because 
of the complicated nature of screening processes in solids. However, it is 
obvious that these interactions are present in real materials. Their strength 
decreases with increasing interatomic distances on the lattice and
they can have important effects on the characteristics of strongly
correlated electron systems. Hence, it is a challenging task to
incorporate and to treat them in an exact way on the level of the
model Hamiltonian. 

The aim of the present paper is two-fold. First, we would like to
extend the previous calculations of Ref.\ \cite{boer1} by choosing a
larger local Hamiltonian which is defined on elementary plaquettes
consisting of four lattice sites of the {\em D} dimensional hypercubic
lattice. Using these local Hamiltonians and a simple, numerically
exact method we have constructed the stability domains for the
 $\eta$-pairing states of momentum $P\!=\!0,\pi$ and for the fully
saturated ferromagnetic state in the parameter space of an extended
Hubbard model with a half-filled band. Given the size of the local
Hamiltonian, the diagonalization is done numerically. The stability
regions are deduced from the equality of the lowest eigenvalue of the
chosen local Hamiltonian and of an upper bound derived appropriately
from the variational principle of quantum mechanics. Second, the
present choice of the larger local Hamiltonian gives us a simple way
to incorporate next-to-nearest neighbor interactions, to treat them
exactly and to investigate their effects on stability.  

The paper is organized as follows: in Sec.\ \ref{sec:method} we
introduce the method. In Sec.\ \ref{sec:hamiltonians} the global
Hamiltonian and the corresponding local Hamiltonian are defined. Sections\
\ref{sec:eta-pairing} and \ref{sec:polarized-FM} contain the stability domains 
in $D\!=\!2,3$ for the $\eta$-pairing states of momentum {\em P} and for the fully 
saturated ferromagnetic case, respectively. Finally a short summary and 
discussion closes the presentation in Sec.\ \ref{sec:conclusions}. 

\section{Method}
\label{sec:method}

Let us consider a model Hamiltonian defined on a discrete
lattice. This (so-called global) Hamiltonian can be decomposed into the sum of 
equivalent local Hamiltonians $h_{\rm cluster}$ defined on identical clusters 
of lattice sites, the union of which covers the lattice. In other words   
\begin{eqnarray}
  \label{eq:Hh}
  H &=& \sum_{\rm all~clusters} h_{\rm cluster} \;.
\end{eqnarray}
If only on-site interactions are considered, each cluster might 
consist of a single lattice point and $h_{\rm cluster}$ is simply a 
Hamiltonian defined on the individual lattice points. If intersite 
interactions are also present, the cluster has to consist of (at least) two 
lattice sites (in case of only nearest neighbor interactions) and the proper 
{\em minimal} local Hamiltonian is actually a Hamiltonian defined on a bond 
joining two (in the simplest case nearest neighboring) sites. Nevertheless, 
the incorporation of longer and longer range interactions, or of more and more 
spatial correlations, requires the enlargement of the minimal cluster
and hence the corresponding local Hamiltonian $h_{\rm cluster}$. In
principle, even infinite range interactions and correlations can be
taken into account, but this would require 
 $h_{\rm cluster}\!=\!H$. The tractable cluster size is, however,
limited by the feasibility of the necessary exact diagonalizations. 

Setting up the local Hamiltonian and choosing a suitable local basis, the 
eigenvalue problem
\begin{eqnarray}
  \label{eq:ev-local}
  h_{\rm cluster}|\phi_{\rm cluster}\rangle &=& \epsilon_{\rm cluster}|
 \phi_{\rm cluster}\rangle
\end{eqnarray}
can be solved exactly. Thus the full spectrum $\epsilon^{i}_{\rm cluster}$
($i\!=\!1,\ldots,{\rm dim}(h_{\rm cluster})$) of $h_{\rm cluster}$ can be
obtained. Since the clusters of decomposition are equivalent, the exact 
ground state energy $E_{\rm GS}$ is bounded from below by the relation 
\begin{eqnarray}
  \label{eq:lower-b}
  E_{\rm lower} &=& N_{\rm cluster} \min_{i} \epsilon^{i}_{\rm cluster} 
\leq E_{\rm GS} \;,
\end{eqnarray}
where the number of clusters on the considered lattice can be written
as $N_{\rm cluster}\!=\!fL$. Here {\em f} represents a simple combinatorical
factor  depending on the dimension and structure of the underlying lattice and
{\em L} stands for the number of lattice sites.

The variational principle of quantum mechanics provides an upper bound for 
$E_{\rm GS}$, namely 
\begin{eqnarray}
  \label{eq:upper-b}
  E_{\rm GS} &\leq& \frac{\langle \Psi_{\rm trial}| H | \Psi_{\rm
      trial} \rangle} {\langle \Psi_{\rm trial}| \Psi_{\rm trial} \rangle} =
  E_{\rm upper} \;, 
\end{eqnarray}
where $|\Psi_{\rm trial} \rangle$ stands for an arbitrary trial wavefunction.
In our case $|\Psi_{\rm trial} \rangle$ is an exact eigenstate of the global
Hamiltonian for a certain set of model parameters. 

Combining now Eqs.\ (\ref{eq:lower-b}) and (\ref{eq:upper-b}) yields 
\begin{eqnarray}
  \label{eq:bound}
  E_{\rm lower} \leq E_{\rm GS} \leq E_{\rm upper} \;.
\end{eqnarray}
Exploiting that both $E_{\rm lower}$ and $E_{\rm upper}$
are analytic functions of the couplings of the global Hamiltonian,
after carefully adjusting the coupling constants one can 
satisfy the equality 
\begin{eqnarray}
  \label{eq:solution}
   E_{\rm lower} &=& E_{\rm upper} \equiv E_{\rm GS} \;, 
\end{eqnarray}
which means that for a certain set of model parameters (in a special 
sector of the ground state phase diagram) 
the exact ground state energy $E_{\rm GS}$ is found. Furthermore, if the
ground state has no degeneracy, the exact ground state $|\Psi_{\rm GS}
\rangle$ of the global Hamiltonian is also
found. In our case the non-degeneracy is provided by the fact that the states
we consider (see Sec.\ \ref{sec:results}) can be built up simply by using
the lowest energy eigenstate of the local Hamiltonian. As an example let us
consider the case of the $D\!=\!1$ dimensional half-filled chain with bonds as
clusters. For certain values of the coupling constants it can be reached that 
the non-degenerate lowest eigenvalue of the local Hamiltonian belongs to the 
parallel orientation of spins included on the bond. Since the bonds are 
equivalent, all the bonds along the chain contain parallel-oriented spins 
with the same local energy for the given set of model parameters. This fact 
yields the long range order of spins (here ferromagnetism) and also the 
non-degeneracy of the global ground state (apart from the spin-degeneracy). 
Similar arguments hold for the non-degeneracy of the ground state in 
higher dimensions, too.

Changing now the trial wavefunction and following the procedure discussed 
above a new ground state in a different region of the phase diagram can be 
found. Furthermore, repeating the method with more and more trial 
wavefunctions a large portion of the ground state phase diagram of the model 
can be explored.
  
\section{Global and local Hamiltonian}
\label{sec:hamiltonians}

Let us define now our global Hamiltonian on a {\em D} dimensional hypercubic
lattice ($D\!>\!1$) in the following form
\begin{eqnarray}
  \label{eq:globH-1}
  H_{\rm glob} &=& U \hat{U} \,+\, \sum_{i,j} \, \left[ -t_{ij} \hat{t}_{ij} 
\,+\, X_{ij} \hat{X}_{ij} \,+\, V_{ij} \hat{V}_{ij} \,+\, Y_{ij} \hat{Y}_{ij} 
\,+\, J_{ij} \hat{J}_{ij} \right] \:-\: \mu \hat{\mu} \;,
\end{eqnarray}
where
\begin{eqnarray*}
  \hat{U} &=& \sum_{i} \: n_{i\uparrow}n_{i\downarrow} \nonumber \\
  \hat{t}_{ij} &=& \sum_{\sigma} \: c_{i\sigma}^{\dagger} c_{j\sigma} 
\nonumber \\
  \hat{X}_{ij} &=& \sum_{\sigma} \: c_{i\sigma}^{\dagger} c_{j\sigma} \,
(n_{i,-\sigma} + n_{j,-\sigma}) \nonumber \\
  \hat{V}_{ij} &=& \frac{1}{2} \: \sum_{\sigma,\sigma'} \: n_{i\sigma}
n_{j\sigma'} \nonumber \\
  \hat{Y}_{ij} &=& c_{i\uparrow}^{\dagger} c_{i\downarrow}^{\dagger} 
c_{j\downarrow} c_{j\uparrow} \nonumber \\
  \hat{J}_{ij} &=& \frac{1}{2} \, \Delta_{ij}^{XY} \left( S^{+}_{i}S^{-}_{j} +
S^{+}_{j}S^{-}_{i} \right)  \:+\: \Delta_{ij}^{Z} \, S^{z}_{i}S^{z}_{j} 
\nonumber \\
  \hat{\mu} &=& \sum_{i,\sigma} \: n_{i\sigma} \;.   
\end{eqnarray*}

Here the fermion operators $c_{i\sigma}^{\dagger}$ ($c_{i\sigma}$) create 
(annihilate) electrons with spin $\sigma$ in the single tight-binding Wannier 
orbital associated with site $i$. $n_{i\sigma}$ is the particle number 
operator of electrons with spin $\sigma$, and 
$n_{i}\!=\!n_{i\uparrow}+n_{i\downarrow}$. Furthermore, the spin
operators are given by  
$S^{+}_{i}\!=\!c_{i\uparrow}^{\dagger}c_{i\downarrow}$,
$S^{-}_{i}\!=\!c_{i\downarrow}^{\dagger}c_{i\uparrow}$ and 
$S^{z}_{i}\!=\!\frac{1}{2}(n_{i\uparrow}-n_{i\downarrow})$. The above 
Hamiltonian contains a term corresponding to the familiar Hubbard term
of doubly occupied sites ($U$), the hopping of a single
electron (characterized by $t_{ij}$), a density dependent (or correlated) 
hopping term ($X_{ij}$), an intersite density-density type interaction of 
electrons ($V_{ij}$), a term describing the hopping of electron pairs 
($Y_{ij}$), for $\Delta_{ij}^{XY}\!=\!\Delta_{ij}^{Z}\!=\!1$ a Heisenberg-type 
exchange interaction of spins ($J_{ij}$) and a chemical potential term 
($\mu$). We note that $\Delta_{ij}^{XY}\!=\!0$ and $\Delta_{ij}^{Z}\!=\!1$ 
represents an Ising-type coupling of electron spins, while the case of
 $\Delta_{ij}^{XY}\!=\!1$ and $\Delta_{ij}^{Z}\!=\!0$ corresponds to an
XY-type interaction. $J_{ij}\!>\!0$ ($J_{ij}\!<\!0$) means antiferromagnetic
(ferromagnetic) type of exchange. The relevance of the
present model for real materials is discussed e.g. in 
Refs.\ \cite{amadon,kollar}.   

In the rest of the paper we investigate only couplings between on-site, 
nearest-neighbor and next-to-nearest neighbor electrons. Incorporating this
restriction into Eq.\ (\ref{eq:globH-1}) and using the convention
 $A_{l}\!=\!A_{ij}$ ($l\!=\!1,2$ for $i$, $j$ being nearest and next-to-nearest
neighboring sites, respectively) for the intersite couplings, one can
rewrite the global Hamiltonian as
\begin{eqnarray}
\label{eq:globH-2}
H_{\rm glob} &=& U \: \sum_{i=1}^{L} \: \left( n_{i\uparrow}-\frac{1}{2} \right) 
\left( n_{i\downarrow}-\frac{1}{2} \right) \\
&+& \sum_{l=1}^{2} \sum_{\langle ij \rangle_{l}} \left\{ \frac{1}{2}
\sum_{\sigma} \left[
X_{l}(n_{i,-\sigma} + n_{j,-\sigma}) - t_{l} \right] ( c_{i\sigma}^{\dagger}
c_{j\sigma} + c_{j\sigma}^{\dagger} c_{i\sigma} ) \:+\: 
\frac{1}{2} \: V_{l} (n_{i}-1)(n_{j}-1) \right. \nonumber \\ 
~~~~~~~~~~~~&+& \left. \frac{1}{2} \; Y_{l} ( c_{i\uparrow}^{\dagger} 
c_{i\downarrow}^{\dagger} c_{j\downarrow} c_{j\uparrow} + 
c_{j\uparrow}^{\dagger} c_{j\downarrow}^{\dagger} c_{i\downarrow} 
c_{i\uparrow} ) \:+\:  
\frac{1}{2} J_{\rm xy}^{(l)} (S^{+}_{i}S^{-}_{j} +
S^{+}_{j}S^{-}_{i}) \:+\: J_{\rm z}^{(l)} S^{z}_{i}S^{z}_{j} \right\} \:-\: 
\mu \hat{\mu}  \:-\: E_{0} \nonumber \;
\end{eqnarray}
Here $\sum_{\langle ij \rangle_{l}}$ means a summation over nearest neighboring
($l\!=\!1$) and next-to-nearest neighboring ($l\!=\!2$) sites. The new
notations $J_{\rm xy}^{(l)}\!=\!\Delta_{l}^{XY}\!J_{l}$ and 
 $J_{\rm z}^{(l)}\!=\!\Delta_{l}^{Z}\!J_{l}$ are also used. In the special 
case of $\Delta_{l}^{XY}\!=\!\Delta_{l}^{Z}\!=\!1$, however, the
 $J_{l} \equiv J_{\rm xy}^{(l)}\!=\!J_{\rm z}^{(l)}$ notation will be
kept for the sake of clarity. $E_{0}$ represents a numerical constant
which shifts the zero point of the energy scale. We remark that during the
reformulation of Eq.\ (\ref{eq:globH-1}) to Eq.\ (\ref{eq:globH-2})
the chemical potential is also shifted by some constant.     

\begin{figure}[htbp]
  \begin{center}
    \leavevmode
    \epsfig{file=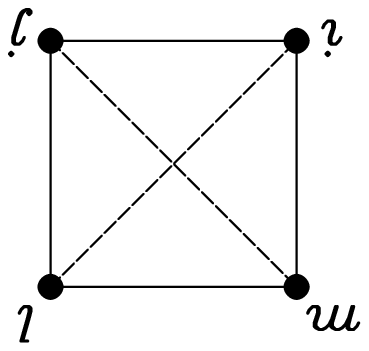, width=5cm, bbllx=460, bblly=230,
      bburx=340, bbury=110}
    \caption{An elementary plaquette of the $D\!=\!2$ dimensional square
      lattice. Solid lines represent different types of nearest
      neighbor couplings while dashed lines symbolize the
      next-to-nearest neighbor ones. The local Hamiltonian is defined on this 
      cluster.}
    \label{fig:fig1}
  \end{center}
\end{figure}

Because we have only nearest and next-to-nearest neighbor interactions,
the fully symmetric minimal clusters are the two-dimensional elementary 
plaquettes of the $D\!=\!2$ dimensional square lattice, as depicted in 
Fig.\ \ref{fig:fig1}. All $D \!>\! 2$ dimensional hypercubic lattices can be 
covered with the elementary plaquettes, however, in those cases  the {\em D} 
dimensional hypercubes would be the fully symmetric minimal clusters. 

Following the method of Sec.\ \ref{sec:method} the global Hamiltonian
can be rewritten in terms of such plaquettes ($\sum_{[i,j,l,m]}$ means a
summation over the plaquettes) as 
\begin{eqnarray}
H_{\rm glob} &=& \sum_{[i,j,l,m]} h_{ijlm}
\end{eqnarray}
with
\begin{eqnarray}
\label{localH}
h_{ijlm} &=& \frac{U}{z_{2}} \sum_{\alpha \in {\cal A}_{0}} 
\left( n_{\alpha \uparrow}-\frac{1}{2} \right) \left( n_{\alpha \downarrow}-
\frac{1}{2} \right) \nonumber \\
&+& \frac{X_{1}}{4f_{1}} \! \sum_{(\alpha,\beta) \in {\cal A}_{N}} 
\sum_{\sigma} \left( c_{\alpha,\sigma}^{\dagger} c_{\beta,\sigma} +
  c_{\beta,\sigma}^{\dagger}c_{\alpha,\sigma} \right) \left(
  n_{\alpha,-\sigma} + n_{\beta,-\sigma} \right) \nonumber \\
&+& \frac{X_{2}}{f_{2}} \! \sum_{(\alpha,\beta) \in {\cal A}_{NN}} 
\sum_{\sigma} \left( c_{\alpha,\sigma}^{\dagger} c_{\beta,\sigma} +
  c_{\beta,\sigma}^{\dagger}c_{\alpha,\sigma} \right) \left(
  n_{\alpha,-\sigma} + n_{\beta,-\sigma} \right) \nonumber \\
&-& \frac{t_{1}}{4f_{1}} \! \sum_{(\alpha,\beta) \in {\cal A}_{N}} 
\sum_{\sigma} \left( c_{\alpha,\sigma}^{\dagger} c_{\beta,\sigma} +
  c_{\beta,\sigma}^{\dagger}c_{\alpha,\sigma} \right) 
\:-\: \frac{t_{2}}{f_{2}} \! \sum_{(\alpha,\beta) \in {\cal A}_{NN}} 
\sum_{\sigma} \; \left( c_{\alpha,\sigma}^{\dagger} c_{\beta,\sigma} +
  c_{\beta,\sigma}^{\dagger}c_{\alpha,\sigma} \right)  \nonumber \\ 
&+& \frac{V_{1}}{4f_{1}} \! \sum_{(\alpha,\beta) \in {\cal A}_{N}}  
\left( n_{\alpha}-1 \right) \left( n_{\beta}-1 \right)   
\:+\: \frac{V_{2}}{f_{2}} \! \sum_{(\alpha,\beta) \in {\cal A}_{NN}}  
\left( n_{\alpha}-1 \right) \left( n_{\beta}-1 \right) \nonumber \\
&+& \frac{Y_{1}}{4f_{1}} \! \sum_{(\alpha,\beta) \in {\cal A}_{N}} 
\left( c_{\alpha\uparrow}^{\dagger}c_{\alpha\downarrow}^{\dagger}
c_{\beta\downarrow}c_{\beta\uparrow} +
c_{\beta\uparrow}^{\dagger}c_{\beta\downarrow}^{\dagger}c_{\alpha\downarrow}
c_{\alpha\uparrow} \right)   
\:+\: \frac{Y_{2}}{f_{2}} \! \sum_{(\alpha,\beta) \in {\cal A}_{NN}} 
\left( c_{\alpha\uparrow}^{\dagger}c_{\alpha\downarrow}^{\dagger}
c_{\beta\downarrow}c_{\beta\uparrow} +
c_{\beta\uparrow}^{\dagger}c_{\beta\downarrow}^{\dagger}c_{\alpha\downarrow}
c_{\alpha\uparrow} \right) \nonumber \\  
&+& \frac{J^{(1)}_{\rm xy}}{4f_{1}} \! \sum_{(\alpha,\beta) \in {\cal A}_{N}} 
\left( S^{+}_{\alpha}S^{-}_{\beta} + S^{+}_{\beta}S^{-}_{\alpha} \right)
\:+\: \frac{J^{(2)}_{\rm xy}}{f_{2}} \! \sum_{(\alpha,\beta) \in {\cal A}_{NN}} 
\left( S^{+}_{\alpha}S^{-}_{\beta} + S^{+}_{\beta}S^{-}_{\alpha} \right)
 \nonumber \\
&+& \frac{J^{(1)}_{\rm z}}{2f_{1}} \! \sum_{(\alpha,\beta) \in {\cal A}_{N}}
S^{z}_{\alpha}S^{z}_{\beta}
\:+\:  J^{(2)}_{\rm z} \!\!\! \sum_{(\alpha,\beta) \in {\cal A}_{NN}}
S^{z}_{\alpha}S^{z}_{\beta} 
\:-\: \frac{\mu}{z_{2}} \sum_{\alpha \in {\cal A}_{0}}  n_{\alpha} 
\end{eqnarray}
Here $f_{1}\!=\!z_{2}/z_{1}$ and $f_{2}\!=\!2$ are numerical constants 
($z_{1}\!=\!2D$ and $z_{2}\!=\!4 {D \choose 2}$  are  the number of nearest 
and next-to-nearest neighboring sites, respectively, on the {\em D}
dimensional hypercubic lattice). They are needed to avoid double or
higher counting of intersite interactions during the plaquette
summation. Furthermore, $t_{1}$, $X_{1}$, $V_{1}$, $Y_{1}$ are the values of
single electron hopping, correlated hopping, density-density type interaction
and pair-hopping between nearest neighboring sites, respectively, while
$t_{2}$, $X_{2}$, $V_{2}$, $Y_{2}$ indicate the analogous processes between
next-to-nearest neighboring sites. $U$ is the Hubbard interaction which can 
either be positive or negative in our model. To mimic real
systems, however, it should be repulsive. In addition, spin interactions with
exchange couplings $J^{(1)}_{\rm xy}$, $J^{(2)}_{\rm xy}$, $J^{(1)}_{\rm z}$ 
and $J^{(2)}_{\rm z}$ are included on the plaquette. Further notations: 
 ${\cal A}_{0}\!=\!\{i,j,l,m\}$ means the set of individual lattice
points, ${\cal A}_{N}\!=\!\{(i,j),(j,l),(l,m),(m,i)\}$ represents the
set of nearest neighboring sites and 
 ${\cal A}_{NN}\!=\!\{(i,l),(j,m)\}$ indicates the set of
next-to-nearest neighboring sites on the plaquette (see Fig.\
\ref{fig:fig1}). The possibility of anisotropies can be naturally
incorporated into the local (and also global) Hamiltonian via the
non-equivalence of the orthogonal directions of the plaquette. The
effects, however, are not discussed in the present paper.

To apply Eq.\ (\ref{eq:solution}) the connection between the number of
lattice sites {\em L} and the number of clusters $N_{\rm cluster}$  (here
plaquettes) is also needed. Combinatoric considerations give the following 
simple result for this quantity on a {\em D} dimensional hypercubic lattice,
\begin{eqnarray}
  \label{eq:plaqnum}
  N_{\rm cluster} &=& \frac{1}{2}D(D-1)L \;.
\end{eqnarray}

\section{Results}
\label{sec:results}

As an illustration of the method described in Sec.\ \ref{sec:method} we 
consider a few physically interesting states, the $\eta$-pairing
states of momentum {\em P} (Sec.\ \ref{sec:eta-pairing}) which show 
off-diagonal long range order and are hence superconducting, and the
fully saturated ferromagnetic state (Sec.\ \ref{sec:polarized-FM}). We
determine under what conditions these states are the ground states of
the global Hamiltonian $H_{\rm glob}$. All the calculations
presented are done at half filling. Except the $\eta$-pairing state of
momentum $P\!=\!0$, it is not possible to express the results in a compact
analytic form as it has been done earlier in Ref.\ \cite{boer1}. 
Therefore, and for the sake of visualization, we have restricted
ourselves to special cuts of the parameter space in illustrating the
effects of the larger local Hamiltonians and of next-to-nearest neighbor 
couplings. The cuts are chosen in such a way that the corresponding
ground state phase diagrams can easily be compared with the previously
published rigorous results of Strack {\em et al.} \cite{strack2}, 
de Boer {\em et al.} \cite{boer1,boer2} and Montorsi {\em et al.} 
\cite{montorsi}. In principle, one can also investigate the role of
each nearest and next-to-nearest neighbor interactions separately, one
by one, using our method.

\subsection{$\eta$-pairing states of  momentum {\em P}}
\label{sec:eta-pairing}

The definition of the $\eta$-pairing operator of momentum {\em P} is given by 
the relation 
\begin{eqnarray}
  \label{eq:eta-oper}
\eta_{\rm P}^{\dagger} &=& \sum_{j=1}^{L} \; e^{iPj} c_{j\downarrow}^{\dagger}
c_{j\uparrow}^{\dagger} \;.
\end{eqnarray}
Using this operator an $\eta$-pairing state of momentum {\em P} and of 
pairs {\em N} can be constructed as
\begin{eqnarray}
  \label{eq:eta-state}
  |\Psi_{\eta}({\rm N,P}) \rangle &=& K \left( \eta_{\rm P}^{\dagger}
  \right)^{N} |0\rangle  \;,
\end{eqnarray}
where $K\!=\!\left[ \frac{(L-N)!}{L!~N!} \right]^{\frac{1}{2}}$ is a
normalization factor. For further details about the $\eta$-pairing states the
reader is referred to the literature \cite{yang1,yang2,shen,boer2,montorsi}.   
Since we would like the $\eta$-pairing states to be the ground state of our
model, it is instructive to calculate the commutator of the
 $\eta$-operator with the global Hamiltonian $H_{\rm glob}$. One finds that
\begin{eqnarray}
\left[ H , \eta_{\rm P}^{\dagger} \right] &=& \sum_{k=1}^{2} \left[ \;
 \frac{1}{2} \; (X_{k}-t_{k}) 
\: \sum_{\langle jl \rangle_{k}} \left( e^{iPj} + e^{iPl}
\right) ( c_{j\downarrow}^{\dagger} c_{l\uparrow}^{\dagger} \;+\;  
c_{l\downarrow}^{\dagger} c_{j\uparrow}^{\dagger} ) \right. \nonumber \\
&&~~~~ + \left. \frac{1}{2} \; X_{k} \sum_{\langle jl \rangle_{k}} 
\left( e^{iPj} - e^{iPl} \right) \left[ ( n_{l\uparrow} - n_{j\downarrow} )
 c_{l\downarrow}^{\dagger}c_{j\uparrow}^{\dagger} \:+\: ( n_{l\downarrow} - 
n_{j\uparrow}  ) \; c_{j\downarrow}^{\dagger} c_{l\uparrow}^{\dagger} \right] 
\right. \nonumber \\
&&~~~~ + \left. \sum_{\langle jl \rangle_{k}} \left\{ \left( \frac{1}{2} \; 
Y_{k} e^{iPj} - U_{k} e^{iPl} \right) (n_{j}-1) \; c_{l\uparrow}^{\dagger} 
c_{l\downarrow}^{\dagger} \right. \right. \nonumber \\
&&~~~~~~~~~~~~~ + \left. \left.  \left( \frac{1}{2} \; Y_{k} e^{iPl}    
- U_{k} e^{iPj} \right) (n_{l}-1) \; c_{j\uparrow}^{\dagger} 
c_{j\downarrow}^{\dagger} \right\} \right] \:-\: 2 \mu \eta_{\rm
P}^{\dagger} \;.
\end{eqnarray}
Calculating now the quantity 
 $\left[ H_{\rm glob} , (\eta_{\rm P}^{\dagger})^{N} \right] |0\rangle$,
one can easily deduce the parameters where the $\eta$-pairing
states of momentum {\em P} are the ground state of the starting model;
for momentum $P\!=\!0$ one arrives at the requirements $X_{i}\!=\!t_{i}$ and
 $Y_{i}\!=\!2V_{i}$ ($i\!=\!1,2$), while for momentum $P\!=\!\pi$ the
conditions $X_{2}\!=\!t_{2}$ and $Y_{i}\!=\!(-1)^{i}2V_{i}$
($i\!=\!1,2$) must be satisfied. One would also look for
 $\eta$-pairing states of momentum $P \neq 0$ or $P \neq \pi$. These, however,
represent the ground states of the global Hamiltonian in Eq.\
(\ref{eq:globH-1}), where $X_{i}\!=\!t_{i}$ ($i\!=\!1,2$), $U \le
-4t_{1}$ and all the other interaction constants are zero
\cite{aligia,schadschneider}.

Using the $\eta$-pairing states as trial wavefunctions, the upper
bound in the thermodynamic limit for the ground state energy per lattice site
{\em L} is
\begin{eqnarray}
  \label{eq:eta-upper}
  \frac{E_{\rm upper}^{\eta_{\rm P}}}{L} &=&
  \frac{1}{4}(U+z_{1}V_{1}+z_{2}V_{2}) \:+\:
  \frac{1}{2}n(\frac{1}{2}n-1)\sum_{l=1}^{2}  z_{l}V_{l} \:+\:
  \frac{1}{4}n(1-\frac{1}{2}n) \sum_{l=1}^{2}  z_{l}Y_{l}\cos^{l}{P} \:-\: \mu
  n \; .
\end{eqnarray}
Exploiting the constraints between the amplitudes of pair-hopping
 $Y_{i}$ and that of density-density type interaction $V_{i}$ one gets
for the half filled  case ($n\!=\!1$) that 
\begin{eqnarray}
  \label{eq:eta-half-filling}
  E_{\rm upper}^{\eta_{\rm P}} &=& \frac{1}{4} L \left( U \;+\; \sum_{l=1}^{2}
    z_{l} V_{l} \right) \;-\; \mu L \;.
\end{eqnarray}

Despite the fact, that the upper bounds for the
 $\eta$-pairing states of momentum $P\!=\!0$ and $P\!=\!\pi$ are
identical, there is a characteristic difference between the two sets
of wavefunctions; the state $|\Psi_{\eta}({\rm N,P\!=\!\pi}) \rangle$
remains an exact eigenstate of the global Hamiltonian $H_{\rm glob}$ even if 
 $X_{1} \neq t_{1}$.   

\begin{figure}[htbp]
  \begin{center}
    \leavevmode
    \epsfig{file=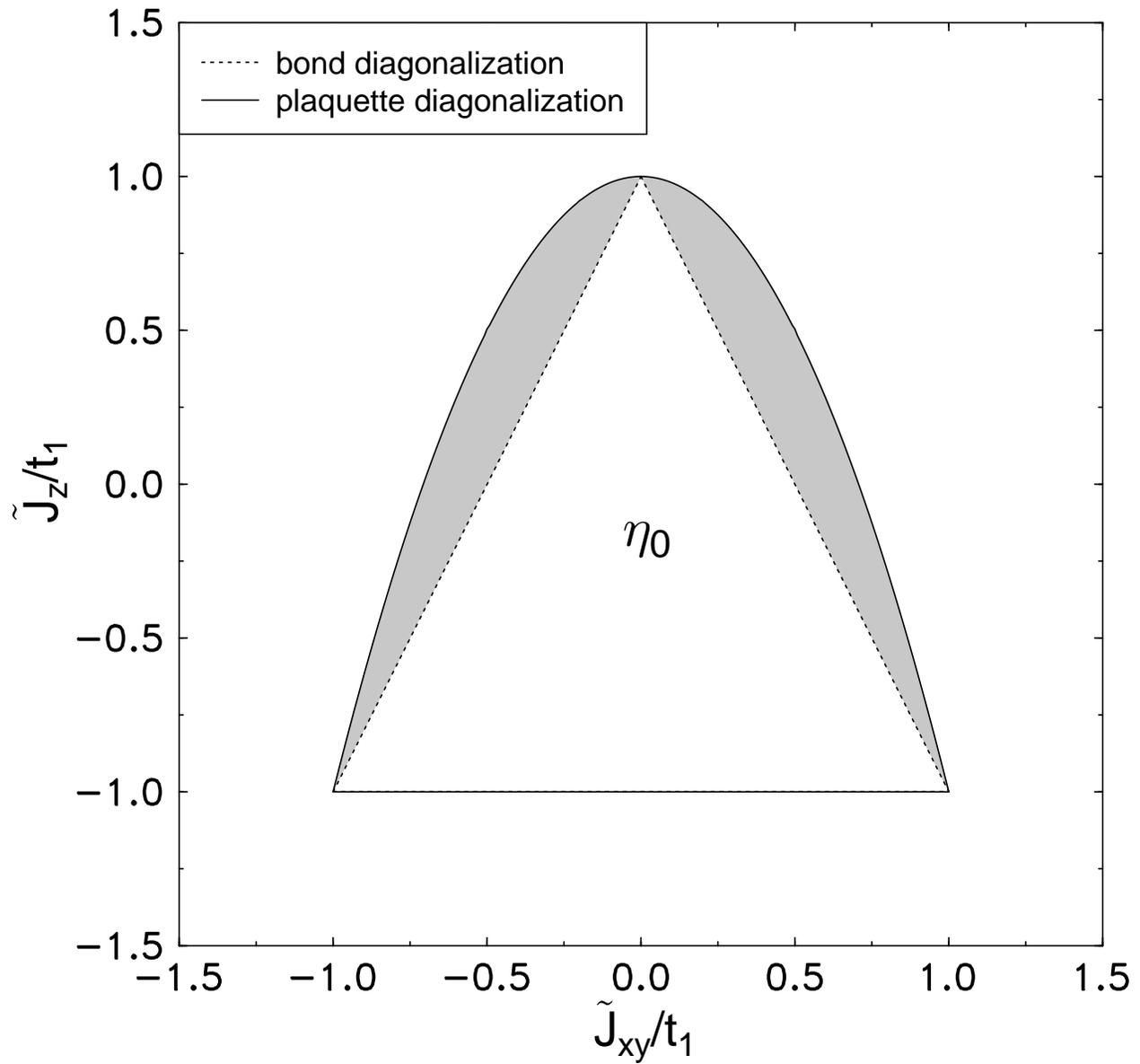, width=18cm, bbllx=560, bblly=460,
      bburx=90, bbury=20}
    \caption{Exact stability region of $\eta$-pairing state of
      momentum $P\!=\!0$ ($\eta_{0}$) at half filling. The shaded region
      represents the enlargement of the stability domain due to the
      choice of local Hamiltonian defined on elementary plaquettes
      (see \protect{Fig.\ \ref{fig:fig1}}) of the lattice. All the
      next-to-nearest neighbor interactions are kept zero.}
    \label{fig:fig2}
  \end{center}
\end{figure}

As mentioned earlier, it is possible to include more spatial
correlations using local Hamiltonians defined on larger clusters of
the lattice. This leads to the extension of the stability of the chosen
state and means an improvement on the phase boundaries. For the 
 $\eta$-pairing state of momentum $P\!=\!0$ this effect is
depicted in Fig.\ \ref{fig:fig2}, where the 
 $\tilde{J}_{\rm xy}$-$\tilde{J}_{\rm z}$ cut of the coupling
constants' space is chosen. The inner triangle corresponds to the
stability domain of the $\eta_{0}$-state determined by the OGS method
of de Boer {\em et al.} \cite{boer1} using bond Hamiltonians with
purely nearest neighbor interactions. The shaded regions give the
improvements of the boundaries applying our method with plaquette
Hamiltonians containing only nearest neighbor interactions. The axes 
 $\tilde{J}_{\rm a}\!=\!J_{\rm a}^{(1)}A(U,V_{1})$ (${\rm a\!=\!xy,z}$)
represent the rescaled values of nearest neighbor 
exchange interactions with the scaling factor of 
\begin{eqnarray}
  \label{eq:eta-scale}
  A(U,V_{1}) &=& \left[2 \, \left|
      \frac{2U}{z_{1}} \,+\, V_{1} \, \right| \,
  \right]^{-1}  \;.
\end{eqnarray}
From Fig.\ \ref{fig:fig2} one can immediately read off the stability criteria
for the $\eta_{0}$-state to be the ground state of $H_{\rm glob}$.
In the absence of next-to-nearest neighbor interactions
(i.e. $t_{2}\!=\!X_{2}\!=\!V_{2}\!=\!Y_{2}\!=\!J_{2}\!=\!0$) for $V_{1}
\leq 0$ values of nearest neighbor density-density type interaction
\begin{eqnarray}
  \label{eq:exact-criteria}
   V_{1} &~\leq~& 0 \nonumber \\
  -1 &~\leq~& \frac{\tilde{J}_{\rm z}}{t_{1}} ~\leq~ -2 
\left( \frac{\tilde{J}_{\rm xy}}{t_{1}} \right)^{\!\!2} + 1 \;. 
\end{eqnarray}
This is a considerable improvement over the $V_{1} \leq 0$,
$-1 \leq \tilde{J}_{\rm z}/t_{1} \leq -2 \, 
|\tilde{J}_{\rm xy}/t_{1}| + 1$ criteria derived following 
Ref.\ \cite{boer1} using bond Hamiltonians. 

If now next-to-nearest neighbor interactions are turn on, i.e. nonzero
values of $V_{2}$ (and hence $Y_{2}$) are considered, the criteria of 
Eq.\ (\ref{eq:exact-criteria}) remains valid in an improved form. In this
case there is a need for the proper change of the scaling factor $A$, 
which now depends also explicitly on next-to-nearest neighbor
interactions. The new value of the scaling factor is 
\begin{eqnarray}
  \label{eq:nn-scale}
    A(U,V_{1},V_{2}) &=& \left[2 \, \frac{z_{2}}{z_{1}} \, \left|
      \frac{2U}{z_{2}}  \;+\; V_{1} \;+\; V_{2} \, \right| \,
  \right]^{-1}  \;.
\end{eqnarray}

\begin{figure}[b]
  \begin{center}
    \leavevmode
    \epsfig{file=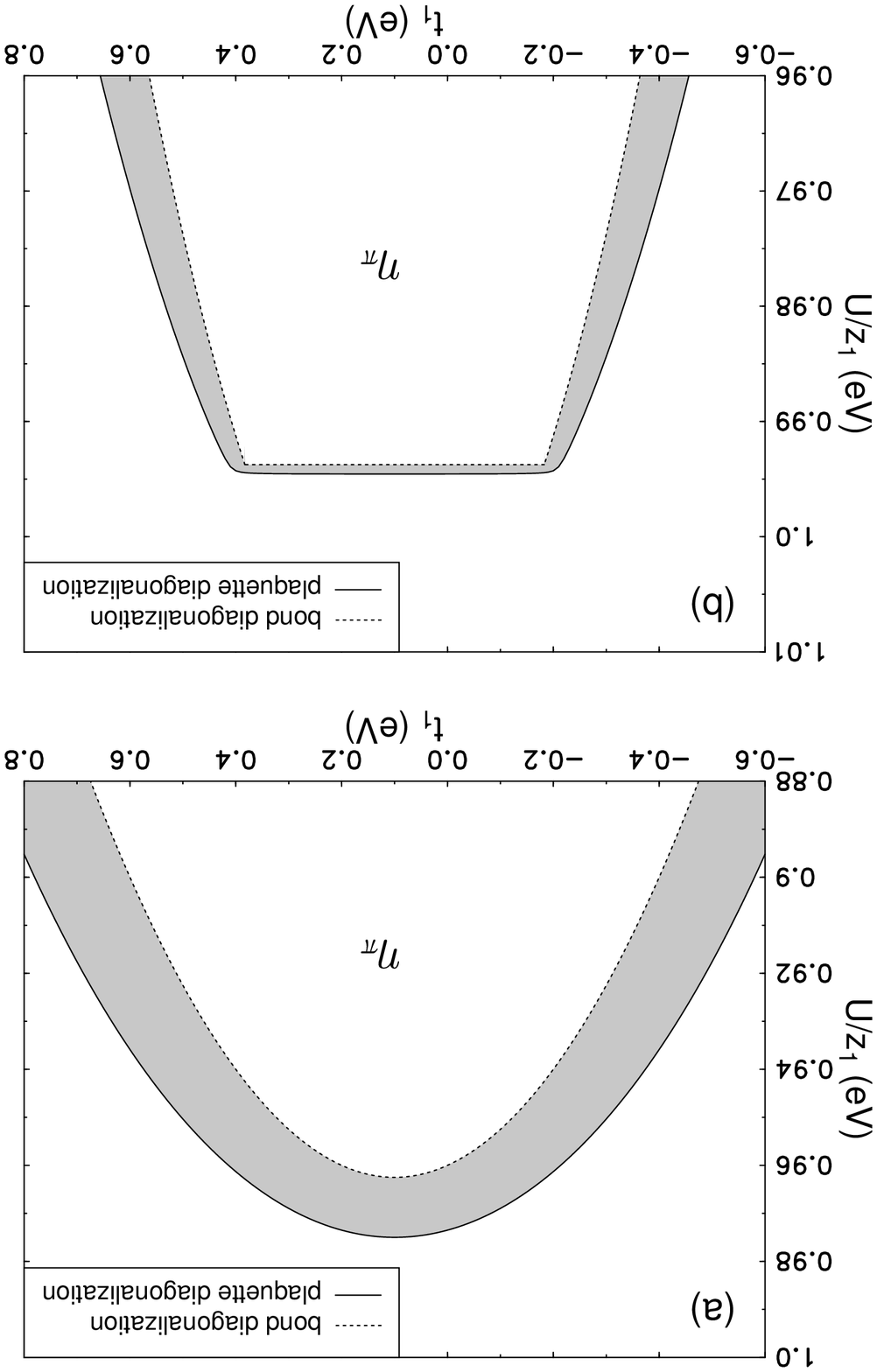, width=16cm, bbllx=590, bblly=700,
      bburx=80, bbury=20}
    \caption{Exact stability domains of $\eta$-pairing state of
      momentum $P\!=\!\pi$ ($\eta_{\pi}$) at half filling for two
      different sets of nearest neighbor couplings in the absence 
      of next-to-nearest neighbor interactions [plot (a): 
      $X_{1}\!=\!0.1~eV$, $V_{1}\!=\!-2~eV$ and
      $J_{1}\!=\!0.05~eV$; plot (b): $X_{1}\!=\!0.1~eV$, $V_{1}\!=\!-2~eV$,
      $J_{\rm xy}^{(1)}\!=\!-0.02~eV$ and $J_{\rm z}^{(1)}\!=\!-0.015~eV$]. The
      shaded regions represent the enlargement of the
      stability domain due to the choice of local Hamiltonian defined
      on plaquettes instead of bonds.}
    \label{fig:fig3}
  \end{center}
\end{figure}

In real systems the $X_{1}\!=\!t_{1}$ requirement does not hold in general.
However, the strengths of the corresponding two interactions are of the same
magnitude. For $X_{1} \neq t_{1}$ only the $\eta$-pairing state of
momentum $P\!=\!\pi$ represents an exact eigenstate of the model
Hamiltonian $H_{\rm glob}$. Figure\ \ref{fig:fig3} shows the stability
regions of the $\eta_{\pi}$-state for two different sets of model parameters
expressed in units of {\em eV} in the {\em U}-$t_{1}$ plane. Dotted
lines represent the boundary for stability regions corresponding to
bond Hamiltonians, while solid lines are the boundaries calculated
with plaquette Hamiltonians in the absence of next-to-nearest neighbor 
interactions. The shaded regions clearly show the extension of
stability regimes due to the choice of larger local Hamiltonians. 

\begin{figure}[htbp]
  \begin{center}
    \leavevmode
    \epsfig{file=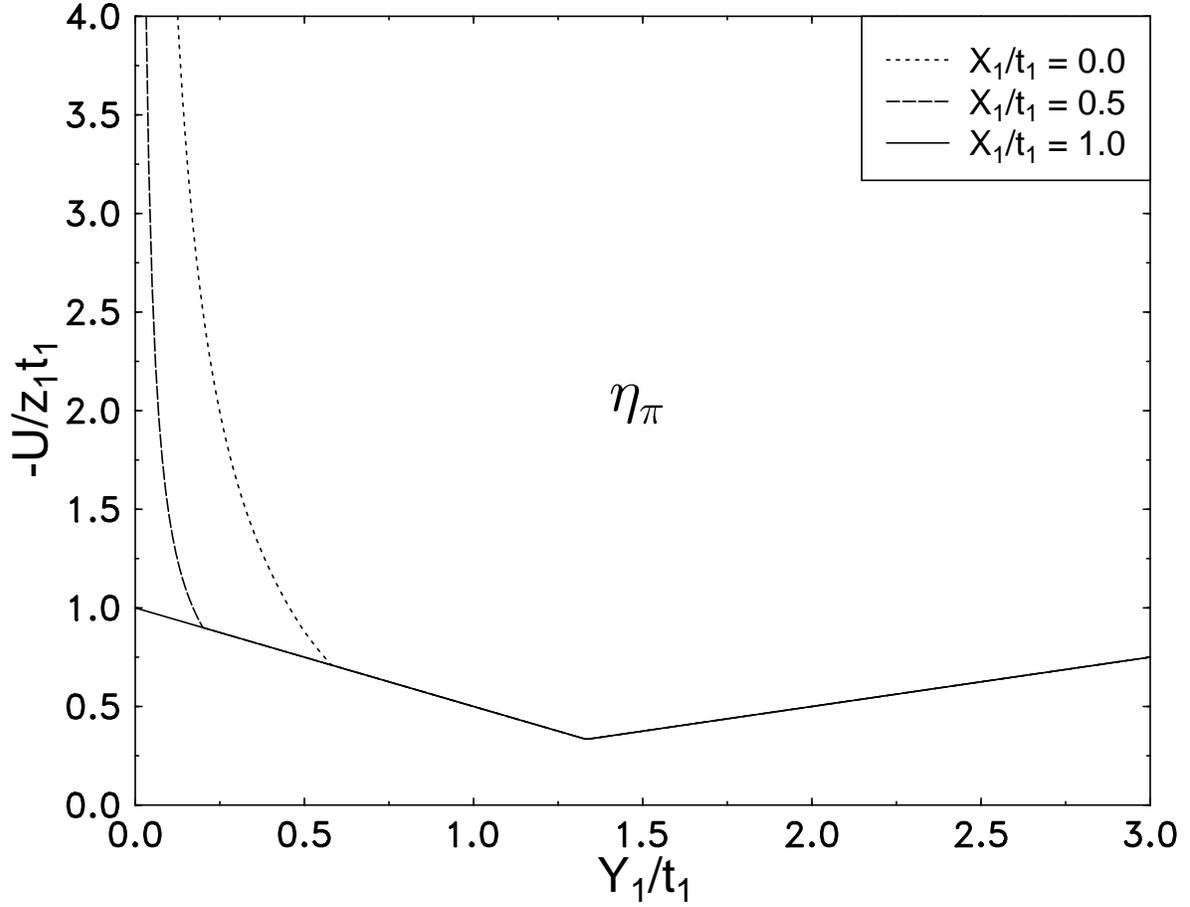, width=17cm, bbllx=540, bblly=380,
      bburx=90, bbury=40}
    \caption{Effects of correlated hopping $X_{1}$ and pair hopping
      $Y_{1}$ on the stability of $\eta$-pairing state of momentum
      $P\!=\!\pi$ ($\eta_{\pi}$). Each next-to-nearest neighbor
      interaction is turned off. We note, that beyond a well-defined
      value of $Y_{1}$ the phase boundaries for the cases 
      $X_{1} \!\neq\! t_{1}$ and $X_{1}\!=\!t_{1}$ coincide.} 
    \label{fig:fig4}
  \end{center}
\end{figure}

\begin{figure}[htbp]
  \begin{center}
    \leavevmode
    \epsfig{file=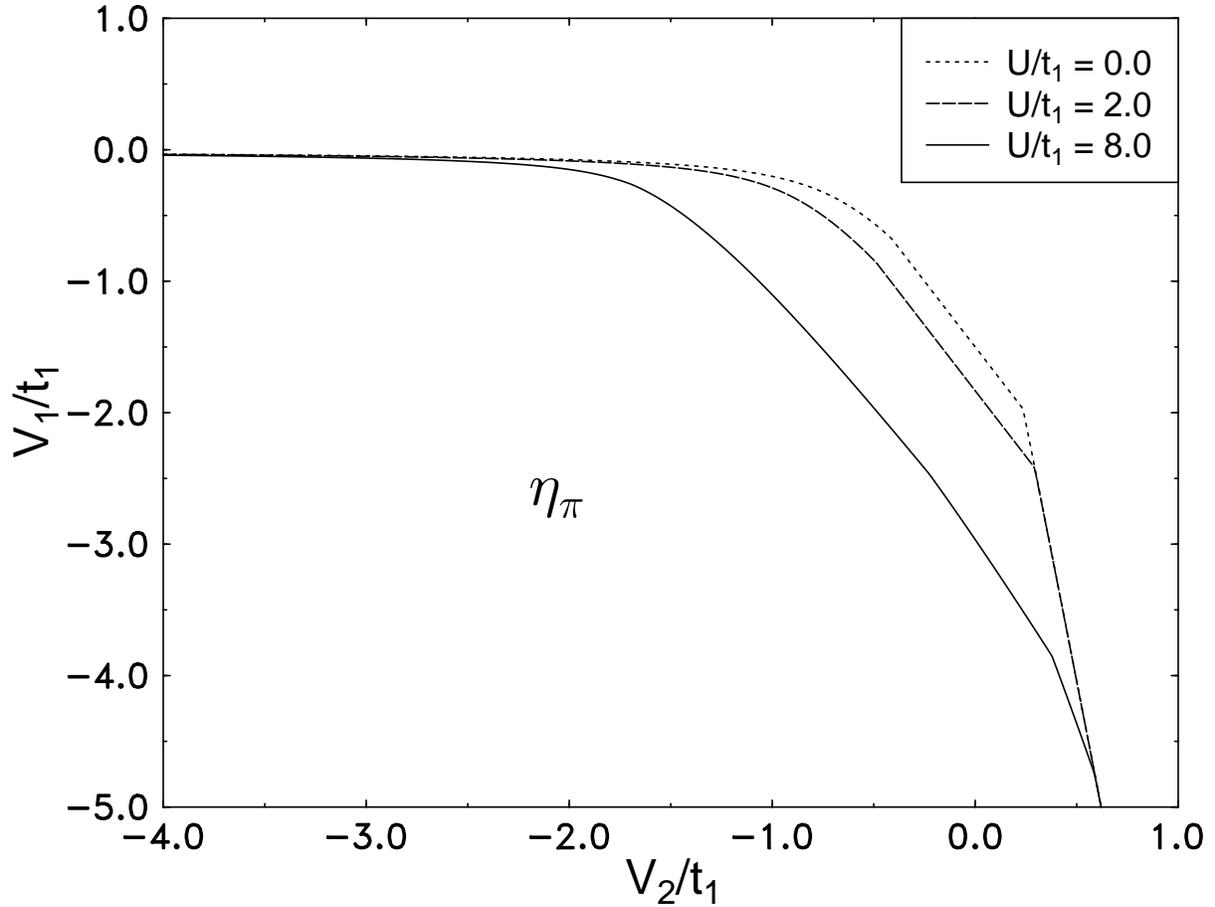, width=17cm, bbllx=540, bblly=380,
      bburx=90, bbury=40}
    \caption{Stability domains of $\eta$-pairing state of momentum
      $P\!=\!\pi$ ($\eta_{\pi}$) in $D\!=\!3$ in the presence of various
      on-site Coulomb repulsions and intersite density-density type
      interactions ($V_{1}$,$V_{2}$). The remaining parameters of Eq.\
      \protect{(\ref{eq:globH-1})} are chosen as follows:
      $t_{2}\!=\!-\frac{1}{4}t_{1}$ and
      $X_{1}\!=\!J_{1}\!=\!J_{2}\!=\!0$. Note, 
      that $X_{2}$,$Y_{1}$ and 
      $Y_{2}$ are uniquely fixed via the rigorous restrictions
      derived earlier in order that Eq.\ \protect{(\ref{eq:eta-state})} be
      an exact eigenstate. In $D\!=\!2$ similar
      stability regions emerge.}
    \label{fig:fig5}
  \end{center}
\end{figure}

We now consider the phase diagrams in the {\em U-$Y_{1}$} (Fig.\
\ref{fig:fig4}) and $V_{2}$-$V_{1}$ (Fig.\ \ref{fig:fig5}) planes for fixed
values of the other interactions. In Fig.\ \ref{fig:fig4} the exchange
interaction has been fixed by the relation $J_{1}\!=\!-2Y_{1}$. This assures
that the global Hamiltonian of Sec.\ \ref{sec:hamiltonians} coincides with 
the model Hamiltonian of Ref.\ \cite{montorsi}, where the authors,
using the method of positive semi-definite operators 
\cite{brandt,strack1,strack2}, derived rigorous bounds for the 
$\eta_{\pi}$-state. Comparing Fig.\ \ref{fig:fig4} with 
Fig.\ \ref{fig:fig1} in Ref.\ \cite{montorsi} two basic differences
can be noticed. First, the boundary of the stability region, even for the
special case of $X_{1}\!=\!t_{1}$, varies with increasing values of nearest
neighbor pair-hopping amplitude $Y_{1}$ and has a maximum at $Y_{1} \!\approx\!
1.33t_{1}$, instead of having a constant value. Second, the value of this
maximum $U_{max} \!\approx\! -0.33z_{1}t_{1}$ is independent of
 $X_{1}$. Furthermore, the stability regions for all values of $X_{1}/t_{1}$ 
derived with the present method are always larger than the corresponding ones
predicted by Montorsi {\em et al.} \cite{montorsi} and de Boer {\em et al.}
\cite{boer1} and exist for all positive values of $Y_{1}$. The
relation $Y_{1}\!=\!-2V_{1}$ is also satisfied because the
 $\eta_{\pi}$-state has to be an exact eigenstate of the
model. Therefore, the positivity of $Y_{1}$ implies that the nearest
neighbor density-density type interaction $V_{1}$ has to be always negative,
i.e. attractive, in order to find an $\eta_{\pi}$ ground state. 

As the $\eta$-pairing states consist of local pairs of
electrons, it is of interest to investigate the effect of on-site Coulomb 
repulsion, characterized by {\em U}, on the stability of these
pairs. To do this the $V_{2}$-$V_{1}$ plane is chosen at various
values of {\em U}. As can be seen in Fig.\ \ref{fig:fig5} the local
pairs are stable even in the presence of relatively large positive values of 
{\em U}. This requires, however, an attraction in the nearest
neighbor density-density interaction channel. It should also be noted that
the same type of interaction between next-to-nearest neighbors, denoted
by $V_{2}$, can either be attractive or moderately repulsive. These
findings lead us to the conclusion that the $\eta$-pairing state of 
momentum $P\!=\!\pi$ remains the ground states of Eq.\ (\ref{eq:globH-1}) even
for positive values of the on-site Coulomb interaction. Hence 
superconductivity can exist in the extended Hubbard model with local
repulsion ($U \!>\! 0$), if a sufficiently strong nearest neighbor
attraction ($V_{1} \!<\! 0$) is present. 

\begin{figure}[htbp]
  \begin{center}
    \leavevmode    
    \epsfig{file=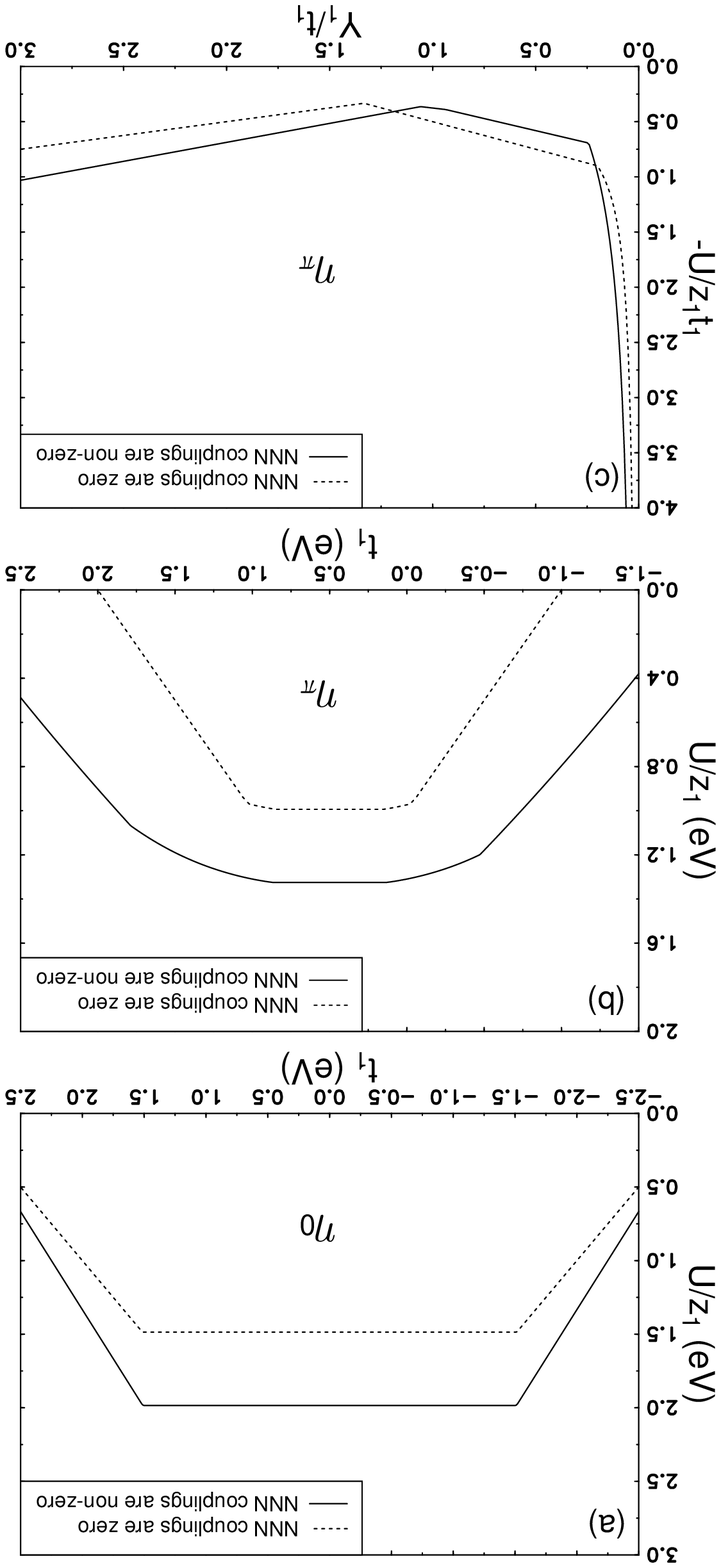, width=12cm, height=24cm, bbllx=540, bblly=730,
      bburx=200, bbury=10}
    \caption{Effects of next-to-nearest neighbor (NNN) couplings on the
      stability of $\eta$-pairing states of momentum {\em P} on a two
      dimensional square lattice. The values of the interaction constants 
      are fixed as follows; plot (a): $P\!=\!0$, $V_{1}\!=\!-3~eV$,
      $J_{\rm xy}^{(1)}\!=\!\frac{1}{40}~eV$, 
      $J_{\rm z}^{(1)}\!=\!\frac{1}{30}~eV$,
      $t_{2}\!=\!\frac{1}{3}t_{1}$, $V_{2}\!=\!\frac{1}{3}V_{1}$  
      and $J_{\rm a}^{(2)}\!=\!\frac{1}{3}J_{\rm a}^{(1)}$ 
      (${\rm a\!=\!xy,z}$); plot (b): $P\!=\!\pi$, $V_{1}\!=\!-2~eV$, 
      $X_{1}\!=\!0.5~eV$, $J_{\rm xy}^{(1)}\!=\!-\frac{1}{50}~eV$,
      $J_{\rm z}^{(1)}\!=\!-\frac{1}{40}~eV$,  
      $t_{2}\!=\!-\frac{1}{3}t_{1}$, $V_{2}\!=\!\frac{1}{3}V_{1}$ and 
      $J_{\rm a}^{(2)}\!=\!\frac{1}{3}J_{\rm a}^{(1)}$ 
      (${\rm a\!=\!xy,z}$); plot (c):
      $P\!=\!\pi$, $X_{1}\!=\!\frac{1}{2}t_{1}$,  
      $J_{1}\!=\!-2Y_{1}$, $t_{2}\!=\!-\frac{1}{5}t_{1}$,
      $Y_{2}\!=\!\frac{1}{8}Y_{1}$ and $J_{2}\!=\!\frac{1}{8}J_{1}$. }
    \label{fig:fig6}
  \end{center}
\end{figure}

The inclusion of next-to-nearest neighbor interactions increases remarkably the
number of model parameters and hence the number of possible cuts of the
parameter space. Therefore, we illustrate only some special, overall
effects of these interactions. In order to model real systems all 
next-to-nearest neighbor interactions are chosen to be smaller in
magnitude than the corresponding nearest neighbor ones. Nevertheless,
the ratio of nearest to next-to-nearest neighbor interactions can be very
different - it depends on the material. In Fig.\ \ref{fig:fig6} three
plots are shown for different values of the couplings. These belong to
the case of $D\!=\!2$. The corresponding $3$-dimensional plots display 
qualitatively the same features, except for the parameters of 
Fig.\ \ref{fig:fig6}c. The quantitative discrepancy between the plots taken in
 $D\!=\!2$ and in $D\!=\!3$ is due to the fact that 
the number of next-to-nearest neighbors $z_{2}$ is much larger in $D\!=\!3$
spatial dimensions than in $D\!=\!2$ on a hypercubic lattice. This
suggests that in the framework of the present model the effects of
next-to-nearest neighbor interactions are stronger in higher dimensions. 

In Fig.\ \ref{fig:fig6}a the stability region 
of the $\eta$-pairing state of momentum $P\!=\!0$ is shown for a
certain set of model parameters in the absence (solid line) and in the
presence (dotted line) of next-to-nearest neighbor couplings. One can
notice the expansion of the stability region due to the presence of 
next-to-nearest neighbor interactions. Since a large value of
 $|t_{1}|$ (in the presence of a fixed value of the nearest neighbor
pair hopping $Y_{1}$) favors the hopping of single electrons over the
hopping of electron pairs, large values of $|t_{1}|$ give rise to
the breaking of local pairs. This means that the number of 
doubly occupied sites is not conserved any longer and, as a
result, the $\eta_{0}$-state ceases to be the ground state of
$H_{\rm glob}$. In Fig.\ \ref{fig:fig6}b the effects of next-to-nearest 
neighbor interactions on the stability of the $\eta_{\pi}$-state are
shown. In contrast to the situation depicted in Fig.\ \ref{fig:fig6}a no
shrinking of the stability domain with increasing $|t_{1}|$ can be
observed. This can be explained with the different internal structure
of the $\eta_{\pi}$ pairs. In Fig.\ \ref{fig:fig6}c we show a
situation, where next-to-nearest neighbor couplings can either
extend or shrink the stability region of the $\eta_{\pi}$-state. As
mentioned earlier, the dimension of the lattice plays a crucial role
here. In $D\!=\!2$ a huge portion of the {\em U}-$Y_{1}$ plane phase 
diagram is occupied by the $\eta_{\pi}$-state for any ratio of 
$X_{1}/t_{1}$. In $D\!=\!3$, however, it was found for a wide 
parameter region that the $\eta_{\pi}$-state is the ground state 
of the model Hamiltonian $H_{\rm glob}$ only for the special 
case of $X_{1}/t_{1}\!=\!1$. 

\subsection{The fully polarized ferromagnetic state}
\label{sec:polarized-FM}

Let us consider now the fully ($z$-)polarized ferromagnetic state
as the trial wave function defined as  
\begin{eqnarray}
  \label{eq:ferro-state}
  |\Psi_{\rm FM} \rangle &=& \prod_{j=1}^{L} c_{j \uparrow}^{\dagger}
  |0\rangle ~=~ \hat{F} |0\rangle \;.
\end{eqnarray}
Calculating the commutator of $\hat{F}$ with $H_{\rm glob}$ one can 
see that this state is an exact eigenstate of the global Hamiltonian for any 
values of the interaction parameters. The trial wavefunction 
$|\Psi_{\rm FM} \rangle$ yields in the thermodynamic limit at 
half-filling the upper bound   
\begin{eqnarray}
  \label{eq:termo-ferro}
  E_{\rm upper}^{\rm FM} &=& -\frac{1}{4} U L \;+\; 
\frac{1}{8} L \sum_{l=1}^{2} z_{l} J_{\rm z}^{(l)} \;-\; \mu L
\end{eqnarray}
for the ground state energy. 

\begin{figure}[htbp]
  \begin{center}
    \leavevmode    
    \epsfig{file=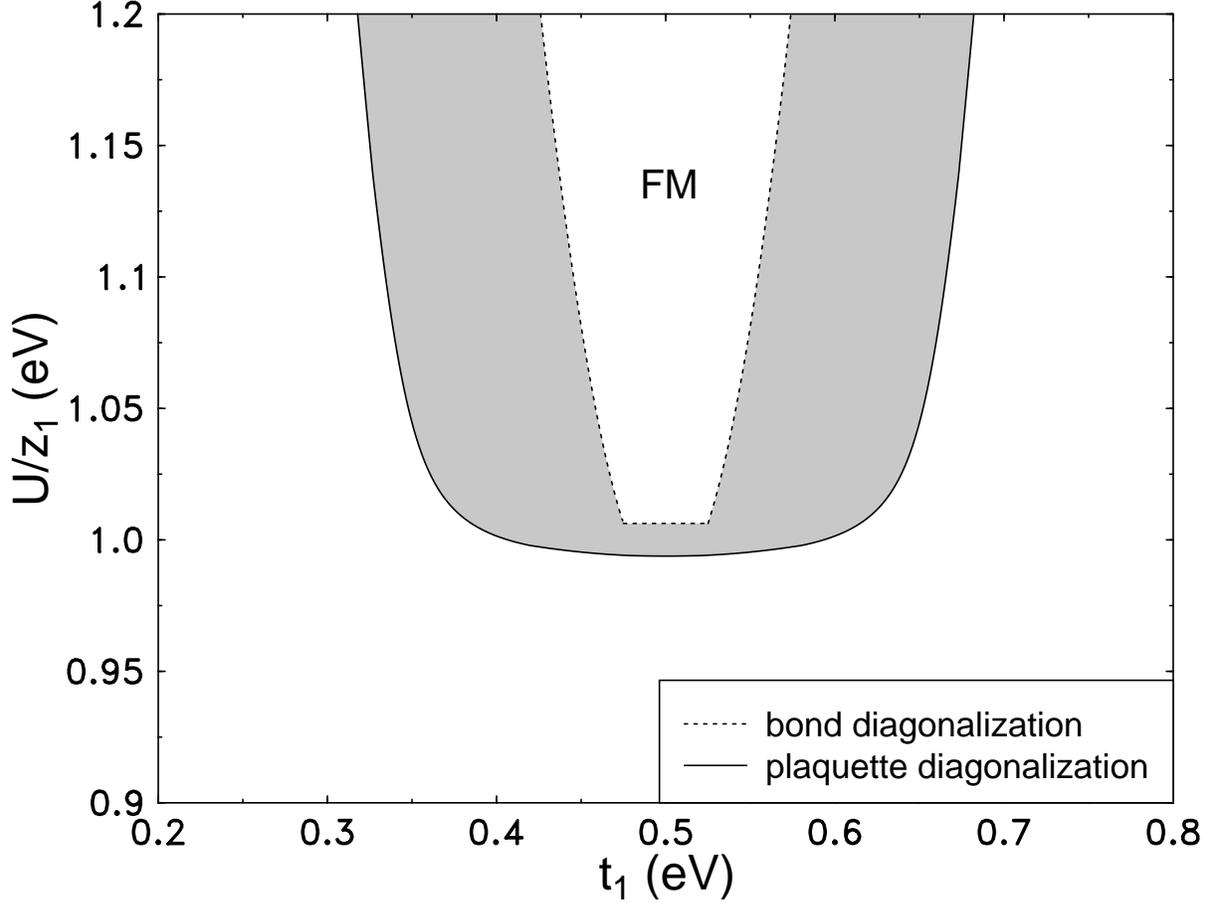, width=17cm, bbllx=540, bblly=380,
      bburx=90, bbury=40}
    \caption{Exact stability region for the fully saturated
      ferromagnetic state (FM) at half filling on a {\em D} dimensional
      hypercubic lattice for a certain set of model
      parameters $X_{1}\!=\!0.5~eV$, $V_{1}\!=\!2~eV$ and
      $Y_{1}\!=\!-J_{1}\!=\!\frac{1}{40}~eV$ in
      the absence of next-to-nearest neighbor interactions. The shaded
      region represents the extension of the stability of the fully saturated
      ferromagnetic state as the ground state of Eq.\
      \protect{(\ref{eq:globH-1})} due to the choice of local Hamiltonian
      defined on elementary plaquettes, instead of bonds, of the lattice.}
    \label{fig:fig7}
  \end{center}
\end{figure}

\begin{figure}[htbp]
  \begin{center}
    \leavevmode
    \epsfig{file=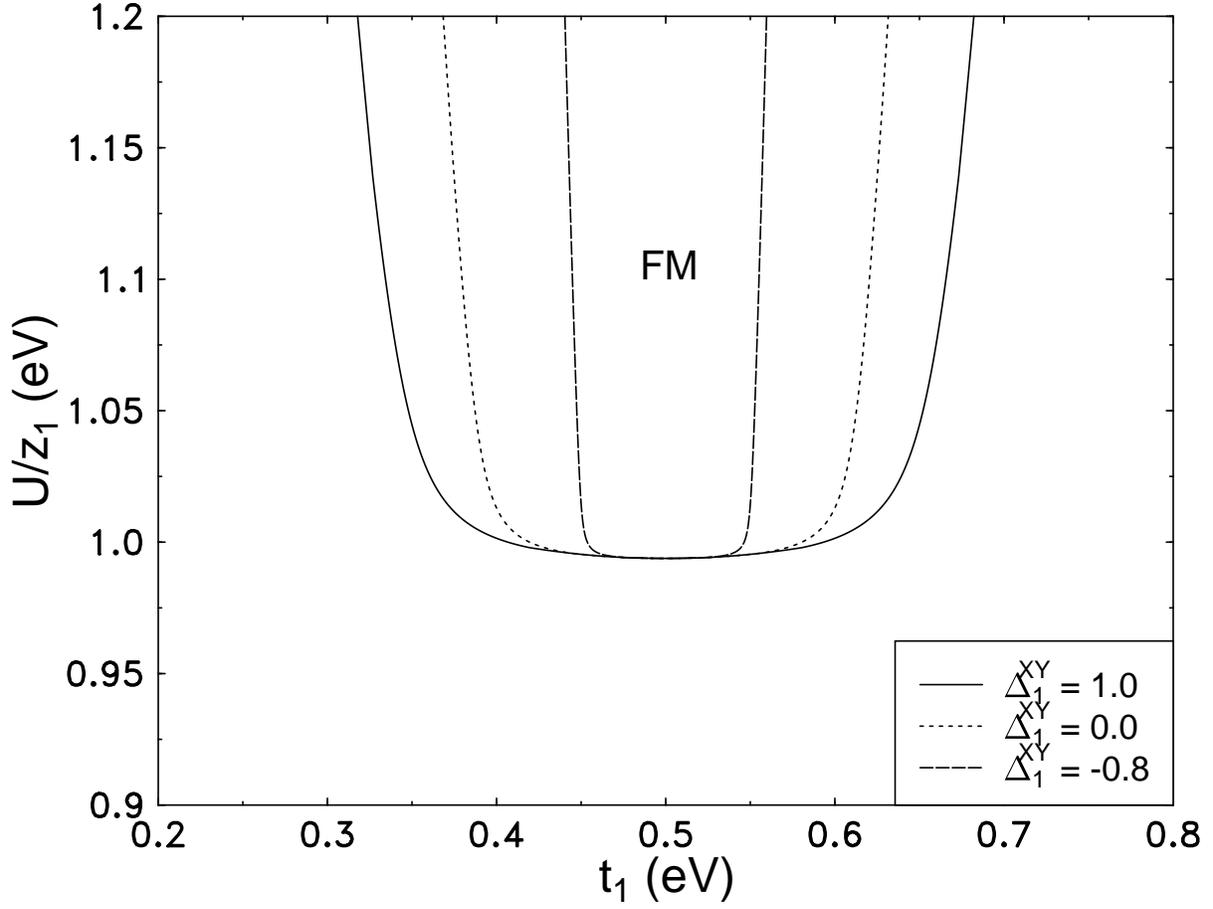, width=17cm, bbllx=540, bblly=380,
      bburx=90, bbury=40}
    \caption{Stability regions of the fully saturated ferromagnetic
      state (FM) for various values of $\Delta_{1}^{XY}$. The
      stability region is maximal at $\Delta_{1}^{XY}\!=\!1$ and vanishes
      for values $\Delta_{1}^{XY} \!<\! -1$. The numerical values of the
      remaining couplings are the same as in \protect{Fig.\ \ref{fig:fig7}}.}
    \label{fig:fig8}
  \end{center}
\end{figure}

\begin{figure}[htbp]
  \begin{center}
    \leavevmode    
    \epsfig{file=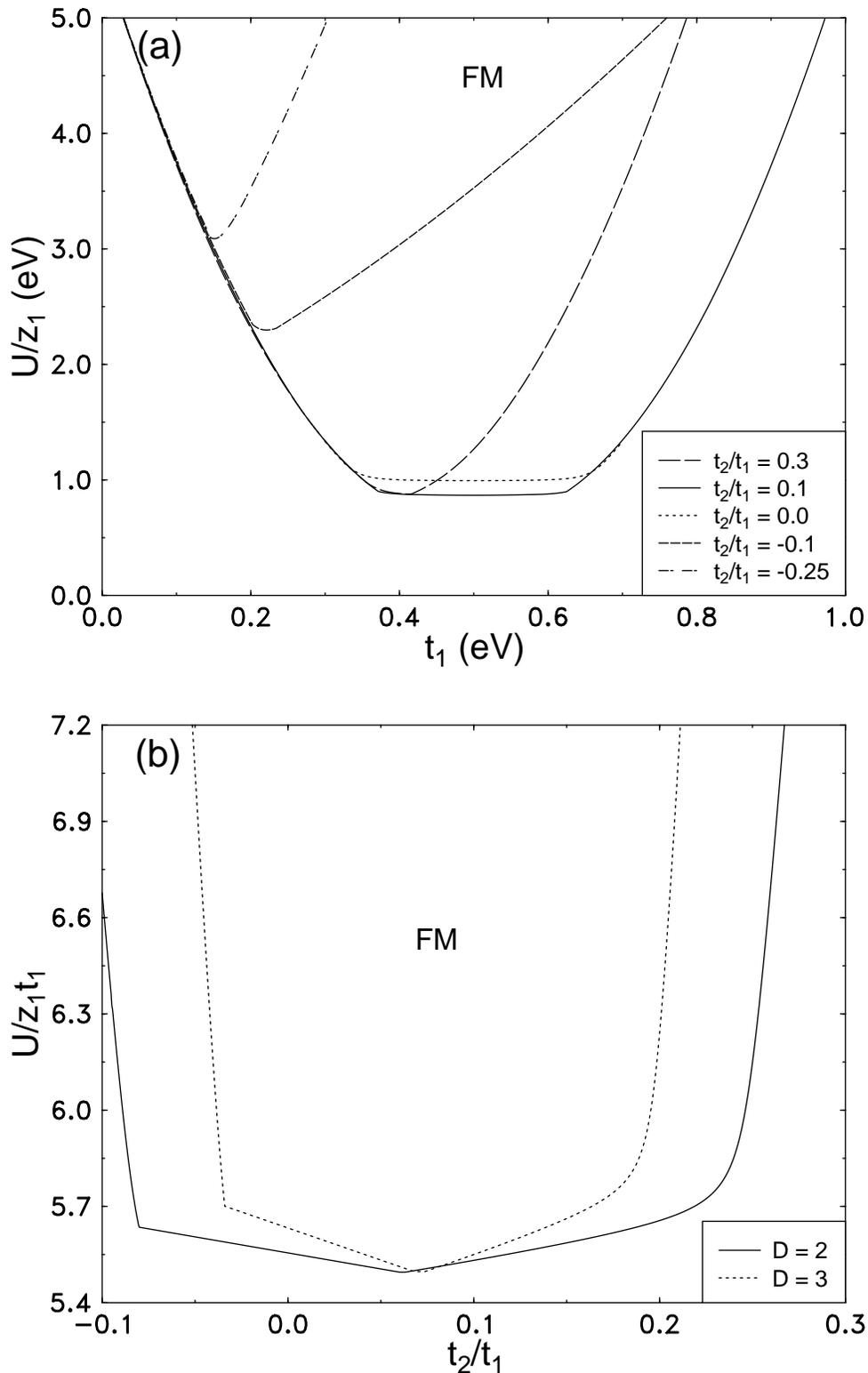, width=14cm, bbllx=530, bblly=700,
      bburx=80, bbury=20}
    \caption{Effects of next-to-nearest neighbor couplings on the
      stability of the fully saturated ferromagnetic state (FM) at
      half filling in the 
      $t_{1}$-{\em  U} and $t_{2}$-{\em  U} planes, plot (a) and (b),
      respectively. Plot (a) shows the phase boundaries
      for various ratios of $t_{2}/t_{1}$ choosing the numerical values of 
      nearest neighbor interactions as in \protect{Fig.\
        \ref{fig:fig7}}. Next-to-nearest
      neighbor interactions for the plot are as follows: $X_{2}\!=\!0.08~eV$,
      $V_{2}\!=\!\frac{1}{8}V_{1}$, $Y_{2}\!=\!\frac{1}{5}Y_{1}$ and
      $J_{2}\!=\!\frac{1}{5}J_{1}$. In plot (b) all the interaction
      constants are
      expressed in units of $t_{1}$ instead of units of {\em eV}. Above each
      line the ground state is the fully polarized ferromagnetic state.}
    \label{fig:fig9}
  \end{center}
\end{figure}

In what follows, we reveal under what conditions Eq.\ (\ref{eq:ferro-state}) 
is the ground state of $H_{\rm glob}$. For the sake of simplicity we 
concentrate on a fixed set of numerical values of nearest neighbor couplings, 
as it has already been estimated by Hubbard \cite{hubbard} for electrons in 
{\em d}-bands of transition metals. The values of next-to-nearest neighbor 
interactions are chosen to be fractions of the corresponding nearest neighbor 
ones. The ratio of nearest to next-to-nearest neighbor couplings is
set to be about {\em 5-8}. We think this range of ratio to be
appropriate for a wide class of materials. Furthermore, this is in
agreement with the work of Appel {\em et al.} \cite{appel} who also
made quantitative predictions regarding the strength of nearest and
next-to-nearest neighbor correlated hopping $X_{1}$ and $X_{2}$,
respectively. Further calculations at different sets of model
parameters have also shown that the phase diagrams plotted in Figs.\
\ref{fig:fig7}, \ref{fig:fig8} and \ref{fig:fig9} are generic. This
suggests that the above choice of model parameters captures the
essential physics.

In Fig.\ \ref{fig:fig7} we present the changes in the stability domain
induced by using plaquette Hamiltonians. For comparison with the
corresponding result of Ref.\ \cite{boer1} with bond Hamiltonians,
the plaquette Hamiltonians contained no next-to-nearest neighbor
interactions. The shaded region shows the enlargement in the stability domain
of the fully saturated ferromagnetic state. As can be seen from the figure,
there is a reasonable extension with respect to $t_{1}$. While in the case of 
bond Hamiltonians the value of the correlated hopping $X_{1}$ should be
very close in magnitude to $t_{1}$ in order to reach the boundary of
the stability region at $U_{\rm min}\!\approx\! 4~eV$, we have a much 
broader region for that using plaquette Hamiltonians. The
broadening implies that the additionally incorporated spatial
correlations really lead to the stabilization of the ordered phase, in
our case the ferromagnetism. 

The global Hamiltonian $H_{\rm glob}$ containing purely nearest
neighbor interactions can be transformed into an effective Heisenberg
model in the large-{\em U} limit at half filling (see e.g. Kollar
{\em et al.}\cite{vollhardt} and references therein) with the effective 
exchange coupling of
\begin{eqnarray}
  \label{eq:t-J-model}
  J_{\rm eff} &=& \frac{t_{1}^{2}}{U} \left(1 -
    \frac{X_{1}}{t_{1}} \right)^{\!2} \,+\, J_{1} \;.
\end{eqnarray}
This favors ferromagnetism for all $J_{\rm eff} \!<\! 0$. Neglecting in
Eq.\ (\ref{eq:globH-1}) all intersite interactions but the nearest neighbor 
exchange interaction, the latter equation suggests a simple perturbative 
criterion for the stability of ferromagnetism in the large-{\em
  U} limit; ferromagnetism is favored over antiferromagnetism for all 
 $U \!>\! U_{\rm c}\!=\!t_{1}^{2}/|J_{1}|$ (note that in 
Eq.\ (\ref{eq:globH-1}) 
 $J_{1}\!=\!-|J_{1}| \!<\! 0$ means the ferromagnetic coupling). It is
known that the OGS approach using bond Hamiltonians gives the criterion 
 $U/z_{1} \!>\! U_{\rm c}$ as the stability requirement in the same regime. 
Therefore it possibly underestimates the stability of the fully polarized 
ferromagnetic state. Taking into account larger local Hamiltonians defined 
on elementary plaquettes of the hypercubic lattice one gets the 
criterion $U/z_{1} \!>\! U_{\rm c}/2$ for 
the stability, which means more extended stability domain in the
large-{\em U} limit. Furthermore, it suggests that (at least in the
large-{\em U} limit) the stability criterion has the form of 
 $U \!>\! \frac{z_{1}}{b} U_{\rm c}$, where $b$ is the function of the
size of the cluster on which the local Hamiltonian is defined. The
possible scaling behavior of the stability criterion and the concrete
form of $b(N_{\rm cluster})$ are discussed elsewhere \cite{szabo2}.

Since the fully polarized ferromagnetic state at half filling is an exact
eigenstate of $H_{\rm glob}$ we have got no {\em a priori}
restrictions for the values of the interaction parameters. The
extensive calculations, however, lead to a simple restriction between 
 $J_{\rm xy}^{(1)}$ and $J_{\rm z}^{(1)}$. In order to have a
ferromagnetic ground state of a model containing spin interactions 
which vary continuously from a Heisenberg-type interaction to a simple 
Ising-type one, the $-1 \!<\! \Delta_{1}^{XY} \!\leq\! 1$ requirement must
hold. This means that the restriction 
\begin{eqnarray}
  \label{eq:restr-FM}
  - |J_{\rm z}^{(1)}| \leq  - J_{\rm xy}^{(1)} < |J_{\rm z}^{(1)}|
\end{eqnarray}
has to be always satisfied. In Fig.\ \ref{fig:fig8} the consequence of
Eq.\ (\ref{eq:restr-FM}) is illustrated in the {\em U}-$t_{1}$ cut of the
parameter space at $\Delta_{1}^{Z}\!=\!1$ and 
 $J_{\rm z}^{(1)}\!=\!-|J_{\rm z}^{(1)}| \!<\! 0$ for various values of
$\Delta_{1}^{XY}$. The size of the stability region is maximal at
$\Delta_{1}^{XY}\!=\!1$ and gradually decreases as $\Delta_{1}^{XY}$ 
reaches $\Delta_{1}^{XY}\!=\!-1$. Any further decrease of $\Delta_{1}^{XY}$ 
yields that our ferromagnetic state which is fully $z$-polarized is no 
longer the ground state of $H_{\rm glob}$; for $J_{\rm
  xy}^{(1)}\!=\!J_{\rm z}^{(1)}$ $H_{\rm glob}$ is SU(2) symmetric
which implies the degeneracy of the
ground state with respect to SU(2) rotations. For anisotropic exchange
couplings favouring the $xy$-plane, the ground state polarization may
still be macroscopic, however, it no longer points in the
$z$-direction. This means that the fully $z$-polarized ferromagnetic trial
wavefunction becomes unstable. We note that Eq.\  (\ref{eq:restr-FM}) must 
hold even in the presence of next-to-nearest neighbor couplings.   

In Fig.\ \ref{fig:fig9} the effects of nonzero next-to-nearest neighbor hopping
$t_{2}$ are illustrated for a fixed set of model parameters, first in the 
{\em U}-$t_{1}$ plane [plot (a), in units of {\em eV}] and second in the {\em
  U}-$t_{2}$ plane [plot (b), in units of $t_{1}$]. In Fig.\ \ref{fig:fig9}a
the dotted line represents the phase boundary in the absence of next-to-nearest
neighbor interactions while solid, long-dashed, dashed and dotted-dashed
lines correspond to phase boundaries in the presence of
next-to-nearest neighbor 
interactions for various values of $t_{2}$. As can be seen from the plot,
next-to-nearest neighbor interactions can help in stabilizing ferromagnetism
for the chosen set of model parameters as long as $t_{2}/t_{1}$ has a
small, positive value. For negative or large positive values of $t_{2}/t_{1}$, 
however, the stability domain reduces significantly and stronger Coulomb
repulsion is needed for the stabilization. This means e.g. that for a
reasonably narrow band ($t_{1} \!\approx\! 0.4~eV$) with the ratio 
of $t_{2}/t_{1}\!=\!0.1$, the required minimal stabilizing Coulomb
repulsion is about $U_{\rm min}\!\approx\! 3~eV$. The same value of
{\em U} at $t_{2}/t_{1}\!=\!-0.25$ is about $U_{\rm min}\!\approx\!
30~eV$ which is a magnitude larger. Nevertheless, the above range of
Coulomb interactions can be considered reasonable for real materials. 

In Fig.\ \ref{fig:fig9}b the effects of change in non-interacting
dispersion due to the inclusion of next-to-nearest neighbor hopping
are shown for a square (solid line) and for a cubic (dotted line)
lattice. The stability domains of ferromagnetism depend on the
dimensionality of the lattice and do not coincide. It is interesting
to note that the most favorable values of $t_{2}/t_{1}$ for which the
Coulomb interaction takes its minimal value $U_{\rm min}$ also depend
on the spatial dimension. The shapes of the stability domains are also of
interest: in a well-defined region of $t_{2}/t_{1}$ $U_{\rm min}$
changes only by a slight amount but as soon as the edges of this
region are reached {\em U} increases drastically. This feature
suggests that inside the stability regions a nonzero next-to-nearest
neighbor hopping via the asymmetric density of states
\cite{hanisch,wahle} helps (in the presence of other next-to-nearest
neighbor interactions) in stabilizing ferromagnetism but outside it
destabilizes ferromagnetic ordering. The edges are determined mainly
by the dispersions (hence the shape of the particular density of
states) and tuned further by other interactions being present in
 $H_{\rm glob}$. 

\begin{figure}[htbp]
  \begin{center}
    \leavevmode    
    \epsfig{file=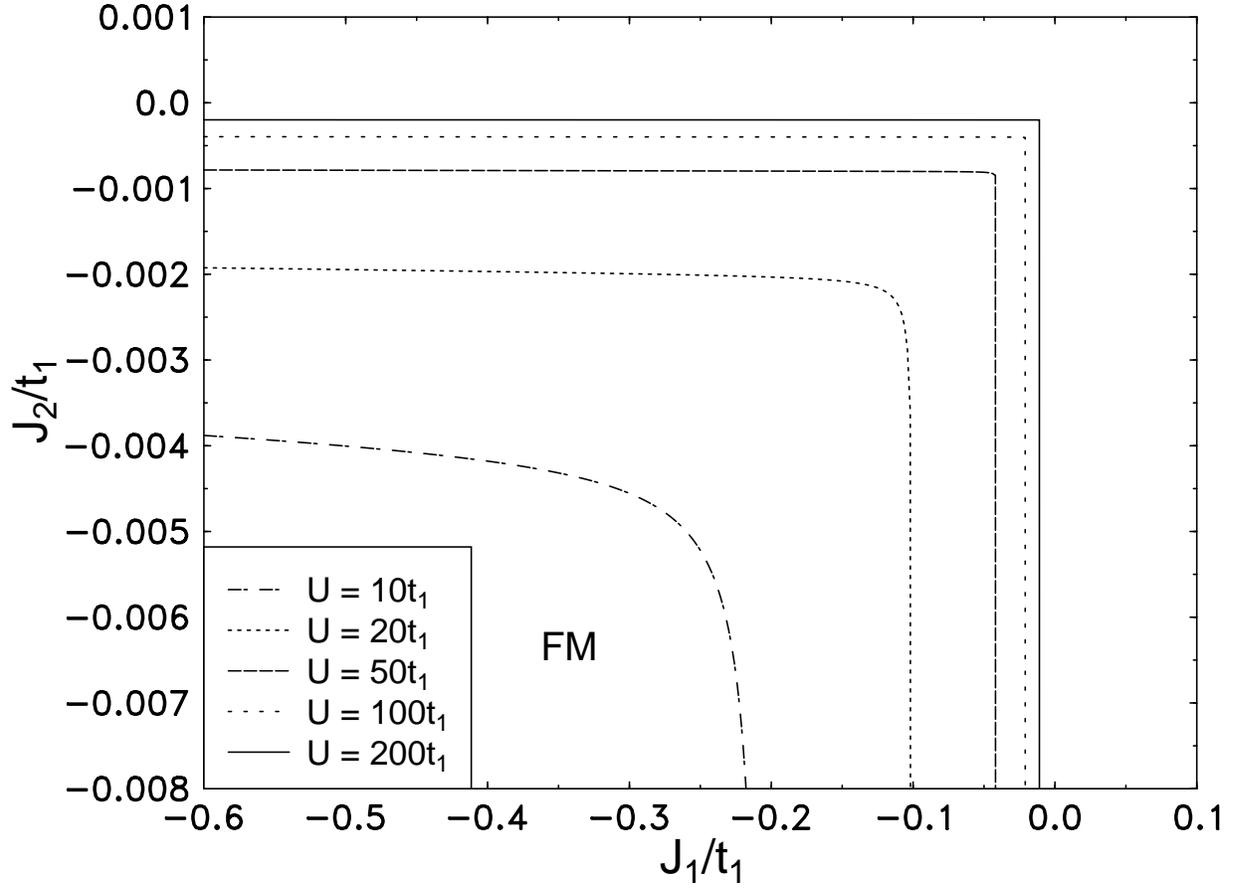, width=17cm, bbllx=560, bblly=380,
      bburx=100, bbury=40}
    \caption{Stability of the fully saturated ferromagnetic state (FM) in the
      presence of nearest ($J_{1}$) and  next-to-nearest ($J_{2}$) neighbor
      exchange coupling at $t_{2}\!=\!-\frac{1}{10}t_{1}$. All the other
      interactions are turned off,
      i.e. $X_{1}\!=\!X_{2}\!=\!V_{1}\!=\!V_{2}\!=\!Y_{1}\!=\!Y_{2}\!=\!0$.} 
    \label{fig:fig10}
  \end{center}
\end{figure}

It is also known that the inclusion of nearest neighbor ferromagnetic exchange 
interaction in the pure Hubbard model favors the parallel ordering of electron 
spins \cite{amadon,strack2,hirsch1}. In Fig.\ \ref{fig:fig10} we
considered the pure Hubbard model supplemented with next-to-nearest
neighbor hopping $t_{2}$ and nearest and next-to-nearest neighbor
exchange interactions $J_{1}$ and $J_{2}$, respectively. All the other
type of couplings are turned off. As can be seen in the figure, the stability
domain of ferromagnetism extends with increasing value of the Coulomb
repulsion and in the limit of $U \!\rightarrow\! \infty $ fills the whole
 $J_{1},J_{2} \!\leq\! 0$ quarter of the phase diagram. The fully polarized
ferromagnetic state remains the ground state of Eq.\ (\ref{eq:globH-1}) for 
finite values of {\em U} only in the presence of finite values of $J_{1}$ and
$J_{2}$. Figure\ \ref{fig:fig10} also shows that the required minimal values 
of $|J_{2}|$ are about an order of magnitude less than the required minimal
values of $|J_{1}|$, and $J_{2}$ should also be ferromagnetic in nature,
i.e. $J_{2} \!<\! 0$.  

\section{Conclusions}
\label{sec:conclusions}

In the present paper we have studied in the thermodynamic limit the
ground state phase diagram of the Hubbard model supplemented by nearest
and next-to-nearest neighbor interactions. The purpose of the study
was to clarify to what extent and in which way the inclusion of additional
spatial correlations changes the stability of physically interesting states,
the $\eta$-pairing state of momentum $P\!=\!0$ and $P\!=\!\pi$, or the fully 
 $z$-polarized ferromagnetic state. The phase boundaries are extracted
from the equality of an upper and a lower bound of the ground state
energy, hence these are exact. The additional spatial correlations are
introduced via the computation of the lower bound on elementary plaquettes,
instead of bonds, of the {\em D} dimensional hypercubic
lattice. Except the case of $\eta$-pairing state of momentum $P\!=\!0$
the exact phase boundaries cannot be given in closed, analytic
forms. Instead, they are shown graphically in special cuts of the parameter
space of the model under study. 

The phase boundaries presented are {\em sufficient} phase boundaries. 
This means that outside the region defined by the exact conditions a 
certain state might remain the ground state of the
model. Diagonalizing local Hamiltonians defined on larger clusters of 
the lattice the phase boundaries 
might be further improved. This improvement with increasing cluster
size addresses a further issue: does the stability domain of a given
ordered phase extend further with taking larger and larger clusters,
or is there a convergence regarding the location of its phase boundary? If the
latter holds, we could determine the phase boundaries with a simple
extrapolation even in the limit of $H\!=\!h_{\rm cluster}$. Based on
preliminary results we believe that the further extension of stability domains
decreases rapidly with increasing cluster size. For instance,
computing the lower bounds of the ground state energy with
diagonalizing local Hamiltonians defined on clusters of 6 lattice
sites, the further expansion of the stability regions is only about a few
percentage, generally 4-5 \% or less. 

Considering the effects of more spatial correlations, which was
equivalent in our case with the choice of plaquette Hamiltonians, we
have improved significantly the previously derived rigorous results
of Refs.\ \cite{strack2,boer1,boer2,montorsi}. This means e.g. for the fully
polarized ferromagnetic state that the minimal value of the Coulomb
repulsion required to stabilize ferromagnetism along reasonable values
of nearest neighbor interactions is predicted to be about 
6-10 {\em eV} (in $D\!=\!3$) in a relatively broad range of nearest
neighbor hopping (see Fig.\ \ref{fig:fig7}).

Another goal of the present study was to determine the overall effects of
next-to-nearest neighbor interactions on the stability domains. The
inclusion of next-to-nearest neighbor interactions can be done
naturally using plaquette Hamiltonians. Up to our knowledge, the
effects of next-to-nearest neighbor interactions, except those of
next-to-nearest neighbor hopping, have not yet been considered
rigorously in the literature. The relative strengths of next-to-nearest
neighbor interactions are much smaller
than that of nearest neighbor interactions, the stability conditions,
however, strongly depend on them. Their effect in various sectors of
the phase diagram is different and they result either an extension or
a shrinking of the stability domain. For instance, taking the
 $\eta$-pairing state of momentum $P\!=\!\pi$ the possible maximal value of
the Coulomb repulsion $U_{\rm max}$, up to which the $\eta_{\pi}$-state
remains the ground state of the extended Hubbard model (i.e. the model has a
superconducting ground state), is increased from 6 {\em eV} to 8-9 {\em eV}
(in $D\!=\!3$) by the inclusion of relatively small next-to-nearest neighbor
interactions (Fig.\ \ref{fig:fig6}b).   

It is also known that next-to-nearest neighbor hopping of single
particles, which is characterized by the hopping amplitude $t_{2}$, is
of importance in real materials. Our results are in good agreement
with this fact. We showed that $t_2$ has a characteristic effect
e.g. on the stability of the fully saturated ferromagnetic state. Even
a small ratio of $t_{2}/t_{1}$, i.e. a small amount of frustration in
the dispersion, introduces a qualitative change into the phase
diagram. The change is mostly a shrinking of the stability domain,
however, for small ratios ($ t_{2}/t_{1} \!\leq\! 0.15$) the presence 
of $t_{2}$ helps in stabilizing ferromagnetism (Fig.\
\ref{fig:fig9}). This is in good agreement with recent DMRG studies
taken on $1$-dimensional triangular lattice \cite{arita}. It is
interesting to note that in our calculations the extension of
ferromagnetic domain occurs always at positive ratios of
$t_{2}/t_{1}$ for fixed values of the other parameters of the model. 

The Hubbard model supplemented only by exchange interactions
$J_{1}$ and $J_{2}$ has also been investigated. Our results are 
in good agreement with Ref.\ \cite{hirsch2}, i.e. the critical values
of nearest and next-to-nearest exchange interactions to give rise to
ferromagnetism approach zero as $U \!\rightarrow\! \infty$ in the case of a
half-filled band in any dimensions. However, at finite values of the
Coloumb repulsion $J_{1}$ and $J_{2}$ should also be finite, if the
ground state is the fully polarized ferromagnetic state. 

In summary, we have established a simple method which allows us to 
incorporate and to treat the effects of next-to-nearest neighbor correlations 
and interactions in an {\em exact} fashion. We showed that the ground 
state of the extended Hubbard  model in the thermodynammic limit at
half filling is superconducting or ferromagnetic, depending on the interaction
strengths. The improved phase boundaries for certain sets of model
parameters have also been constructed.  

\section{Acknowledgements}
\label{sec:acknowledgement}

The author would like to thank Zs. Gul\'acsi and E. M\"uller-Hartmann for
their continuous encouragement during the present work. He also thanks 
A. Schadschneider, F. P\'azm\'andi, G. Uhrig and P. Wurth for valuable
discussions and careful reading of the manuscript. The author is
gratefully acknowledges financial support of Deutscher Akademischer
Austauschdienst (DAAD) and the hospitality of the University of
Cologne, Germany.

\vfill
\newpage

\end{document}